\documentclass[12pt]{article}

\usepackage{times}
\usepackage{verbatim}
\usepackage{amsmath}
\usepackage{hyperref}
\usepackage{graphicx}
\usepackage{amsfonts}
\usepackage{caption}
\usepackage{subcaption}
\usepackage{braket}
\usepackage{slashed}
\usepackage{amsthm}
\usepackage{url}
\newcommand{\di}{\text{d}}

\numberwithin{equation}{section}

\numberwithin{mylemma}{section}

\numberwithin{mycor}{section}
\DeclareMathOperator{\Tr}{Tr}

\newenvironment{sciabstract}{\begin{quote} \bf}
{\end{quote}}

\title{Approximate thermofield dynamics of interacting fermions}

\author
{Edward B. Baker III \\
\\
\normalsize{Institute for Physical Science and Technology, University of Maryland, College Park}
}

\date{\today}

\date{}

\begin{document}

\maketitle

\begin{sciabstract}
We analyze the many-particle Schrodinger equation for fermions in a thermal ensemble by introducing an exponential operator expansion, defined in the context of thermofield dynamics. The expansion is optimized variationally at each time step through changes in the basis of excitations, which leads to a method of generating approximate differential equations to solve the time dependent problem, and can also be used to cool the system in imaginary time.  The method is applied for a specific set of basis transformations and truncation scheme, leading to an explicit set of differential equations that reduce to the Hartree Fock solution in the low temperature limit.  This procedure can also be generalized to include quantum correlation, which will be pursued in a future publication.
\end{sciabstract}

\section{Introduction}

Understanding the properties of many body quantum systems is one of the most important problems in physics and chemistry, and also one of the most challenging.  For large systems, the intractable complexity of an exact solution to the problem is well known\cite{PhysRevLett.94.170201}, specifically for systems of interacting fermions, which suffer from the so-called sign problem\cite{PhysRevB.41.9301}.  There are many approaches to finding approximate solutions in different contexts, and for a variety of systems\cite{fetter2012quantum}\cite{cramer2005essentials}.  In condensed matter theory, an important general strategy is to search for an effective description of the interacting theory in terms of weakly interacting quasi-particles\cite{altland2006condensed}\cite{negele1995quantum}.  This approach has had success for a large variety of systems, including superconductors\cite{PhysRev.108.1175}, low temperature liquids\cite{nozieres1999theory}, crystal lattices\cite{lattices} and plasmas\cite{plasmas}.  There are also important examples of strongly correlated systems that are not as amenable to a description in terms of quasi-particles, for which a variety of techniques are being actively investigated, and have led to some exciting new approaches to the problem\cite{Yanase20031}\cite{doi:10.1021/ct501187j}\cite{doi:10.1142/S0217979201006495}.

In the field of quantum chemistry, there are a number of computational approaches with a long history\cite{HeadGordon:2008ds}\cite{doi:10.1021/cr990029p}, generally leading to a tradeoff between accuracy and scalability. Density functional theory (DFT) is perhaps the most scalable method for large systems of interacting particles, and gives high accuracy in many cases\cite{orio2009density}\cite{dreizler2012density}.  However, DFT must approximate the exchange-correlation energy of the system, which is generally uncontrolled and difficult to quantify, and is known to give inaccurate results for a number of systems\cite{cohen08}. Ab initio methods are generally more accurate and complete descriptions of the system\cite{jensen2007introduction}\cite{hfreview}, but are usually too computationally expensive for large numbers of particles.  In particular, the coupled cluster method is one of the most scalable ab initio techniques\cite{RevModPhys.79.291}, and is similar in form to the approach taken in this paper.  A set of alternative approaches known collectively as Quantum Monte Carlo methods are generally more scalable than ab initio methods but less than DFT, thus allowing for more accurate calculations for reasonably large systems \cite{nightingale1999quantum}\cite{doi:10.1021/acs.jctc.5b00917}\cite{:/content/aip/journal/jcp/142/23/10.1063/1.4921984}.

In this paper, we introduce a technique that is an almagamation of some of the approaches discussed above, but does not fit neatly into any individual category. Our method will use the thermofield formalism, which is an equivalent alternative to the usual density matrix description of quantum statistical ensembles\cite{thermofield}\cite{takahashi1975collective}.   We first show that in a non-interacting theory, the equilibrium in a grand canonical ensemble always takes the form of a thermal coherent state, which is closely related to the usual definition of a Fermionic coherent state.  We then introduce a general expansion whose first term is this coherent state, and formally solve this expansion in terms of solutions of the many-body Schrodinger equation.  The expansion converges poorly for an interacting system, but the introduction of a transformed basis allows for weak effective interactions, in addition to exponential suppression of the excited states at low temperatures. This leads to the possibility of a rapidly converging series if the basis is chosen well, and we develop a method for dynamically optimizing the basis by cooling the system from infinite temperature, using the  imaginary time formalism\cite{:/content/aip/journal/jcp/63/4/10.1063/1.431514}.

Our approach is similar in spirit to the coupled cluster method, expanding the wavefunction as an operator exponential which is formally equivalent to a cumulant expansion\cite{shavitt2009many}.  However, the use of the thermofield formalism makes all of the operators in the exponential commute, which is a useful simplification. Additionally, this approach is inherently intended for thermal ensembles, incorporating some of the advantages of quantum monte carlo techniques, in addition to providing a more general class of calculations than usually possible for ground state methods. 

The method is then applied by truncating the series at the first term, and allowing unitary transformations that mix the two Hilbert spaces, bearing some resemblance to the Hartree-Fock-Bogoliubov method\cite{Goodman_1979}.     It is shown that the symmetries of the problem allow for the choice of a linear combination of operators in the thermofield double, for which the wavefunction is the vacuum of the system. This leads to simplifications in the calculation, from which an explicit set of differential equations for the dynamics is derived.  In the last section, we apply the results to the homogeneous electron gas, deriving a generalization of a familiar result from the Hartree-Fock analysis of this problem.  This technique can be generalized in a number of ways, providing an interesting avenue for future investigation.

\section{Thermal coherent states}\label{sec:two}

We will begin by reviewing the thermofield formalism, which will be used to define a thermal coherent state.  We will then show that these states arise naturally as the equilibrium configuration for a non-interacting system in the grand canonical ensemble. 

Thermofield theory is a way of describing mixed states that is different from the density matrix formalism, but yields equivalent results. Consider a quantum mechanical system defined on a Hilbert space $\mathfrak{H}$ with a fixed number of particles $N$, and Hamiltonian $\hat{H}$ at temperature $T=k_B/\beta$.  The density matrix for the equilibrium configuration of this system is given by 
\begin{equation}
\hat{\rho}_\beta = \frac{1}{Z}e^{-\beta\hat{H}},
\end{equation}
where $Z$ is the partition function for this system.  Operator expectation values in this ensemble are given by 
\begin{equation}\label{eq:expectationvalue}
\braket{\mathcal{O}}_\beta = \Tr(\mathcal{O}\hat{\rho}_\beta).
\end{equation}
In the thermofield formalism, one introduces a fictitious system that includes two copies of the initial Hilbert space $\mathfrak{H}_{tf} = \mathfrak{H}_1 \otimes \mathfrak{H}_2$, called the thermofield double. Operators acting in the first system are of the form $\mathcal{A}_1\otimes I_2$ and will be referred to as $\mathcal{A}_1$, and on the second Hilbert space are defined by the Hermitian conjugate, $\mathcal{A}_2\equiv I_1\otimes \mathcal{A}_2^\dagger$. In the energy eigenbasis, we can define the entangled state
\begin{equation}\label{eq:canonicalpsi}
\ket{\psi_\beta} = \frac{1}{Z^{\frac{1}{2}}}\sum_i e^{-\frac{1}{2}\beta E_i}\ket{E_i,E_i},
\end{equation}
where the state $\ket{E_i,E_j}$ is an energy eigenvector with eigenvalue $E_i$ and $E_j$ in the seperate Hilbert spaces.  A short calculation shows that the operator expectation value in equation \eqref{eq:expectationvalue} is given by
\begin{equation}
\braket{\mathcal{O}}_\beta = \bra{\psi_\beta}\mathcal{O}_1\ket{\psi_\beta},
\end{equation}
which shows the equivalence of this formalism with the usual density matrix approach\cite{thermofield}. 

We now restrict ourselves to a system of fermions in the grand canonical ensemble, with chemical potential $\mu$.  In this case, the Hilbert space $\mathfrak{H}$ includes states with an arbitrary number of particles, and the relevant density matrix is given by 
\begin{equation}\label{eq:granddensitymatrix}
\hat{\rho}_{\mu,\beta} = \frac{1}{\mathcal{Z}}e^{-\beta(\hat{H}-\mu \hat{N})},
\end{equation}
where $\mathcal{Z}$ is the grand partition function.  The expectation value of an operator in this ensemble is again given by $\Tr(\mathcal{O}\hat{\rho}_{\mu,\beta})$.  Let us introduce a complete set of creation operators $\hat{a}_\alpha^\dagger$ acting on $\mathfrak{H}_1$ that satisfy canonical anti-commutation relations
\begin{equation}\label{eq:anticomm}
\{\hat{a}_\alpha,\hat{a}_\beta^\dagger\}=\delta_{\alpha\beta}, \ \ \ \{\hat{a}_\alpha,  \hat{a}_\beta\}=0, \ \ \ \{\hat{a}_\alpha^\dagger,  \hat{a}_\beta^\dagger\}=0.
\end{equation}
 We also introduce a set of corresponding creation operators $\hat{b}_\alpha$ acting on $\mathfrak{H}_2$, where Hermitian conjugation is included, switching  the notation for creation and annihilation operators.  
Define the state 
\begin{equation}\label{eq:grandpsi}
\ket{\psi_{\mu,\beta}} = \frac{1}{\mathcal{Z}^{\frac{1}{2}}} e^{-\frac{1}{2}\beta (\hat{H}_1-\mu\hat{N}_1)}e^{\sum_\alpha  \hat{a}_\alpha^\dagger\hat{b}_\alpha}\ket{0},
\end{equation}
where $\ket{0}$ is the vacuum of the full Hilbert space, and $\hat{N}$ is the number operator. For an observable $\mathcal{O}_1$ that commutes with all operators $\hat{b}_\alpha$, we find the relation 
\begin{equation}\label{eq:grandoperatorequality}
\braket{\mathcal{O}}_{\mu,\beta} = \bra{\psi_{\mu,\beta}}\mathcal{O}_1\ket{\psi_{\mu,\beta}},
\end{equation} 
which shows that the state \eqref{eq:grandpsi} is the equivalent of \eqref{eq:canonicalpsi} for the grand canonical ensemble.

There is a close analogy between equation \eqref{eq:grandoperatorequality} and the Grassmann resolution of the identity used in Fermionic path integrals\cite{negele1995quantum}\cite{opac-b1131978}
 \begin{equation}\label{eq:grassres}
I=\int\prod_{\alpha} \di \xi_{\alpha}^* \di \xi_{\alpha}\, e^{\sum_{\alpha}\xi_\alpha \xi_\alpha^\star}\ket{\xi}\bra{\xi},
\end{equation}
where $\ket{\xi}$ is a Fermionic coherent state
\begin{equation}
\ket{\xi}=e^{\sum_\alpha \hat{a}_\alpha^\dagger\xi_\alpha }\ket{0}.
\end{equation}
 First we note that equation \eqref{eq:granddensitymatrix} can be written in the form
\begin{equation}
\hat{\rho}_{\mu,\beta}=\frac{1}{\mathcal{Z}}e^{-\frac{1}{2}\beta(\hat{H}-\mu\hat{N})}I\,e^{-\frac{1}{2}\beta(\hat{H}-\mu\hat{N})}.
\end{equation} 
Inserting the resolution of the identity \eqref{eq:grassres} into this expression, and taking an operator expectation value yields an inner product with the Grassmann valued wavefunction 
\begin{equation}
\ket{\tilde{\psi}_{\mu,\beta}} = \frac{1}{\mathcal{Z}^{\frac{1}{2}}} e^{-\frac{1}{2}\beta (\hat{H}-\mu\hat{N})}e^{\sum_\alpha  \hat{a}_\alpha^\dagger \xi_\alpha}\ket{0},
\end{equation}
where the inner product includes the Grassmann integrals.  This state is manifestly similar to equation \eqref{eq:grandpsi}, with the Grassmann numbers $\xi_\alpha$ replaced by the operators $\hat{b}_\alpha$.  In this case, the Grassmann integrals in equation \eqref{eq:grassres} impose similar delta functions to the canonical anti-commutation relations of the operators $\hat{b}_\alpha$, and one can readily verify that they yield the same results.  

This analogy motivates us to define a thermal coherent state
\begin{equation}\label{eq:thermcoh}
\ket{b}=e^{\sum_{\alpha}  \hat{a}_\alpha^\dagger\hat{b}_\alpha}\ket{0},
\end{equation}
which satisfies the formal eigenvalue equation
\begin{equation}
\hat{a}_\alpha \ket{b}=\hat{b}_\alpha\ket{b}.
\end{equation}
Let us consider how the state \eqref{eq:grandpsi} evolves in imaginary time $\tau=\frac{1}{2}\beta$ for a non-interacting Hamiltonian 
\begin{equation}\label{eq:nonintHamiltonian}
\hat{H}^{(0)}=\sum_{\alpha\beta}H^{(0)}_{\alpha\beta}\, \hat{a}^\dagger_\alpha \hat{a}_\beta,
\end{equation}
where we will set $\hbar=1$. 
Assume that the state is of the form \eqref{eq:thermcoh} at time $\tau$,  and consider an infinitesimal evolution of the system to time $\tau+\delta \tau$, where the operators $\hat{b}_\alpha(\tau)$ are now taken to be time dependent. Acting with an annihilation operator, we find
\begin{equation}\label{eq:noninteractingevolution}
\hat{a}_\alpha\ket{b(\tau+\delta \tau)}=\left(\hat{b}_\alpha(\tau)- \delta \tau \Big[\hat{a}_\alpha,(\hat{H}^{(0)}-\mu \hat{N})\Big]\right)\ket{b(\tau+\delta \tau)}+O(\delta \tau^2).
\end{equation}
Because the Hamiltonian is non-interacting, the commutator gives only annihilation operators, which also yield eigenvalues up to order $(\delta \tau)^2$. So we see that the state remains in a coherent state whose eigenvalues satisfy the differential equation
\begin{equation}
\partial_\tau \hat{b}_\alpha(\tau) = \mu  \hat{b}_\alpha(\tau)-\sum_{\beta}H_{\alpha\beta}^{(0)}\hat{b}_\beta(\tau).
\end{equation}
This shows that in a non-interacting system, the thermal wave function in the grand canonical ensemble \eqref{eq:grandpsi} is given by a thermal coherent state whose eigenvalues satisfy the above differential equation. 

  Let us choose the operators $\hat{a}_\alpha^\dagger$ to be eigenfunctions of the non-interacting Hamiltonian 
\begin{equation}
[\hat{H}^{(0)}, \hat{a}_\alpha^\dagger]=E^{0}_\alpha\hat{a}_\alpha^\dagger,
\end{equation}
 In this case, the eigenvalues $\hat{b}_\alpha(\tau)$ are given by 
\begin{equation}\label{eq:batau}
\hat{b}_\alpha(\tau)=e^{-\tau(E^{0}_\alpha-\mu)}\hat{b}_\alpha(0).
\end{equation} 
The occupation number of the state $\alpha$ is then given by 
\begin{equation}\label{eq:FermiDirac}
n_\alpha\equiv\braket{\hat{a}^\dagger_\alpha\hat{a}_\alpha}_{\mu,\beta}=\frac{1}{1+e^{\beta(E^{0}_\alpha-\mu)}},
\end{equation}
which is the Fermi-Dirac distribution, as would be expected. The average number of particles and variance are given by  
\begin{align}
&\braket{\hat{N}}_{\mu,\beta}=\sum_\alpha n_\alpha, \nonumber \\
&\braket{(\Delta\hat{N})^2}_{\mu,\beta}=\sum_{\alpha}n_\alpha\Big(1-n_\alpha\Big),\label{eq:variance}
\end{align}
which are also as expected.

 \section{Interacting dynamics}

We will now consider how a two-body interaction changes the results derived above.  We introduce a Hamiltonian of the form 
\begin{equation}\label{eq:fullHamiltonian}
\hat{H}=\hat{H}^{(0)}+\hat{H}^{(1)},
\end{equation}
where $\hat{H}^{(1)}$ is a two-body potential
\begin{equation}\label{eq:Hint}
\hat{H}^{(1)}=\sum_{\alpha\beta\gamma\delta}w_{\alpha\beta\gamma\delta}\hat{a}^\dagger_\alpha\hat{a}^\dagger_\beta\hat{a}_{\gamma}\hat{a}_{\delta},
\end{equation}
and $\hat{H}^{(0)}$ is given by equation \eqref{eq:nonintHamiltonian}.
 In this case, the commutator $[\hat{a}_\alpha,\hat{H}]$ does not consist of terms that only contain annihilation operators.  
Based on equation  \eqref{eq:noninteractingevolution} we can no longer conclude that coherent states are preserved by the dynamics.  This is similar to the situation for Slater determinants, stemming from the intrinsic complexity of interacting systems.

To proceed, we will introduce an operator expansion to include the effects of correlation in a systematic way. First, we introduce some terminology. We say that an operator is \textit{thermofield (tf) normal of order n}, if it is of the form
\begin{align}
\hat{\psi}^{(n)}=\frac{1}{(n!)^2}\sum_{\boldsymbol{\alpha}, \boldsymbol{\beta}}t_{\boldsymbol{\alpha}\boldsymbol{\beta}}\prod_{i=1}^{n} \hat{a}_{\alpha_i}^\dagger\hat{b}_{\beta_i}.
\end{align}
where $\boldsymbol{\alpha}$ and $\boldsymbol{\beta}$ are multi-indices of order n, and $t_{\boldsymbol{\alpha}\boldsymbol{\beta}}$ is a complex valued function of these indices.  These are similar in form to excitation operators in the coupled cluster or configuration interaction expansions, with n-excitation operators corresponding to tf normal operators of order n.  In this case, however, the creation and annihilation operators act in different Hilbert spaces.  We say that a wavefunction is tf normal if it can be written 
\begin{equation}\label{eq:tfnormal}
\ket{\psi}=\left(1+\sum_{n=1}^\infty \hat{\psi}^{(n)}\right)\ket{\Omega},
\end{equation}
where $\hat{\psi}^{(n)}$ are tf normal of order n, and $\ket{\Phi}$ is a reference wavefunction in the full Hilbert space. 

We now rewrite equation \eqref{eq:tfnormal} as an exponential
\begin{equation}\label{eq:tfexp}
\ket{\psi}=\exp\left(\sum_{n=1}^\infty \hat{T}^{(n)}\right)\ket{\Omega},
\end{equation}
where the operators $\hat{T}^{(n)}$ are also tf normal of order n, and the notation is the same as the coupled cluster method. This will be called a thermal cumulant expansion.  Assuming everything converges properly, we can relate the operators in equation \eqref{eq:tfexp} to those in \eqref{eq:tfnormal} through the relation
\begin{equation}\label{eq:chisolution}
\hat{T}^{(n)}=\hat{\psi}^{(n)}-\sum_{p\in \mathcal{P}_n}  \prod_{m=1}^{n-1}\frac{1}{p_m!}\left(\hat{T}^{(m)}\right)^{p_m},
\end{equation}
where $\mathcal{P}_n$ is the set of multi-indices $\{(p_1,\ldots,p_{n-1}):p_m\geq 0, \sum_{m=1}^{n-1} p_mm =n\}$, and $n>0$.  

As discussed in the introduction, this expansion is closely related to the expansion used in the coupled cluster method, and the interpretation is similar. In particular, the expansion has the property of size extensivity, which is a desirable property for a many body ansatz\cite{shavitt2009many}.  The mathematics of this construction is also formally the same as the cumulant expansion, with the corresponding statistical interpretation\cite{doi:10.1143/JPSJ.17.1100}. 

To get some intuition for the expansion in this context, consider expanding the wavefunction $\ket{\psi_{\mu,\beta}}$ of equation \eqref{eq:grandpsi} in this way, for the Hamiltonian \eqref{eq:fullHamiltonian}.  We will again use the imaginary time formalism, setting $\tau=\frac{1}{2}\beta$, and see how the function $\ket{\psi_\mu(\tau)}$ evolves in imaginary time.  The initial conditions are given by 
\begin{align}\label{eq:initcond}
&\hat{T}^{(1)}(0)=\sum_\alpha \hat{a}_\alpha^\dagger  \hat{b}_\alpha \\ 
&\hat{T}^{(n)}(0)=0, \ \ \ \forall  n> 1.\nonumber
\end{align} 
Inserting this into equation \eqref{eq:chisolution} implies the initial conditions
\begin{align}\label{eq:psiinitcond}
\hat{\psi}^{(n)}(0)=\frac{1}{n!}(\psi^{(1)}(0))^n, \ \ \ \forall  n> 1,
\end{align} 
in addition to $\hat{\psi}^{(1)}(0)=\hat{T}^{(1)}(0)$.  The inverse factors of $n!$ generally account for symmetrization in products of the form $\prod_{i=1}^n \hat{b}_{\alpha_i}\hat{a}_{\alpha_i}^\dagger$.  
Similarly, the inverse factor of $p_m!$ in equation \eqref{eq:chisolution} is a combinatorial factor that accounts for permutations among groups of the same size.  In this case, the operators $\hat{\psi}^{(n)}$ will satisfy the modified Schrodinger equation 
\begin{align}\label{eq:schrodevone}
\partial_\tau\hat{\psi}^{(n)}(\tau)=\Big[(\mu\hat{N}-\hat{H}),\hat{\psi}^{(n)}(\tau)\Big].
\end{align} 
We therefore interpret the operators $\hat{T}^{(n)}$ as probing the n-particle correlation of a group of particles, after subtracting off all possible lower order correlations. For example, the function $\hat{T}^{(2)}$ compares the interacting evolution of two particles $\hat{\psi}^{(2)}$ with the uncorrelated evolution of the product state $(\hat{\psi}^{(1)})^2/2$.  In a similar way, equation \eqref{eq:chisolution} subtracts off all possible lower order correlations from the coherent evolution of n particles, isolating the n-particle correlation, which is similar to the statistical interpretation of cumulants. 

This expansion will only converge rapidly if the theory is weakly interacting, because a strongly interacting theory will make the coherent evolution very different from the independent evolution, so the corrections will be large. To see this more explicitly, we again choose the eigenbasis of the non-interacting Hamiltonian $\hat{H}^{(0)}$, where the functions $\hat{T}^{(1)}(\tau)$ are given by
\begin{equation}\label{eq:schrodevtwo}
\hat{T}^{(1)}(\tau)=\sum_\alpha e^{-\tau(E_\alpha^{0}-\mu)}\hat{a}_\alpha^\dagger\hat{b}_\alpha,
\end{equation}
as they were in the non-interacting theory.
Clearly this term will only be a good approximation for the interacting system if the interactions are weak.  This also illustrates why the non-interacting theory only contains the first term in the expansion, because the higher order corrections are identically zero.   

\section{Optimized basis set}

In the discussion so far, the basis of operators has been kept general.  The only restriction on the basis has been the form of the Hamiltonian \eqref{eq:fullHamiltonian}, and the use of the number operator for the grand canonical partition function.  If one considers a more general basis, the Hamiltonian and number operator no longer retain their usual form, but the rest of the results remain unchanged as long as the operators form a complete set that satisfy the anti-commutation relations \eqref{eq:anticomm}, and the Hamiltonian acts in $\mathfrak{H}_1$.  Even more generally, one might consider a basis of operators that have a mixture of Fermionic and Bosonic statistics, similar to the situation for Cooper pairs, or even anyonic commutation relations.  It is also possible to consider basis sets that are not orthonormal in any sense, as in quantum chemistry algorithms that use nonorthogonal basis sets\cite{PhysRevB.59.1743}.   Finally, in the approach taken here, it will be possible to consider transformations that mix the two Hilbert spaces, which will be further explored. As special cases of transformed basis sets, one could consider Bogoliubov transformations\cite{altland2006condensed}\cite{negele1995quantum}, local canonical  transformations\cite{PhysRevB.44.12413}\cite{citeulike:13177435}, or more general nonlinear canonical transformations\cite{BVtwomodes}.   Some general results on canonical transformations are reviewed in appendix \ref{sec:AppA}.

We will develop a method to incrementally optimize the basis set so that the cumulant expansion \eqref{eq:thermcoh} converges as rapidly as possible.  Ideally, the full wavefunction could be accurately approximated by the first term in the infinite sum, giving it the general form of a thermal coherent state.  To achieve this goal, we must introduce a truncation scheme, and then maximize the overlap of the truncated wavefunction with the ``full'' wavefunction, over the set of allowed basis transformations.    The truncation scheme is a choice of terms to keep in the exponential, as considered in the coupled cluster method.  For example, one could keep only operators of first order, conventionally called singles, or also include doubles, or higher order terms.

 The following analysis will apply to real or thermal time evolution, so for convenience we define the modified Hamiltonian 
\begin{align}\label{eq:calH}
\hat{\mathcal{H}}=
\begin{cases}
\hat{H}\ \ \ \ \  \ \ \ \ \ \ \  \ \ \ \ \  \ \ \ \ \ \ real,
\\
-i(\hat{H}-\mu \hat{N}) \ \ \ \  \ thermal,
\end{cases}
\end{align}
where we note that the number operator must be defined in the physical basis of particle excitations, and then transformed to whatever basis is being used.
Let us assume that the basis has been optimized at time $t$, and try to optimize it at time $t+\delta t$. 
  At time $t$, the truncated wavefunction has the form
\begin{equation}
\ket{\psi(t)_s}=e^{\hat{T}_s(t)}\ket{\Omega},
\end{equation}
where the subscript $s$ denotes a truncated (or short) cumulant expansion.  At time $t+\delta t$, the evolved wavefunction will become  
\begin{equation}
\ket{\psi(t+\delta t)}=e^{-i\delta t \hat{\mathcal{H}}}e^{\hat{T}_s(t)}\ket{\Omega}.
\end{equation}
After using the Baker-Campbell-Hausdorff formula, it is clear that the state will no longer be truncated, and will in general have terms of all orders in the exponential. We would like to find a truncated wavefunction $\ket{\upsilon}$ that maximizes the overlap $\lvert\braket{\upsilon|\psi(t+\delta t)}\rvert^2$, within a set $\Upsilon$ of wavefunctions that have a constant magnitude, so that
\begin{equation}\label{eq:variational}
\ket{\psi(t+\delta t)_s}=\underset{\upsilon\in\Upsilon_n}{\operatorname{argmax}}\lvert\braket{\upsilon|\psi(t+\delta t)}\rvert^2.
\end{equation}
We do not demand the wavefunction to be normalized because we will take advantage of the freedom to simplify the equations.  

It is desirable to turn equation \eqref{eq:variational} into a differential equation using the calculus of variations.  To do this, one can use Lagrange multipliers, and vary the state $\bra{\upsilon}$ indepenently of $\ket{\upsilon}$ to obtain
\begin{equation}\label{eq:variation1}
\braket{\delta\upsilon|\psi}\braket{\psi|\upsilon}=\lambda\braket{\delta\upsilon|\upsilon},
\end{equation}
where we suppress the time dependence, and the subscript $s$, for convenience. To obtain a dynamic equation, we take a derivative of this equation, set $\ket{\upsilon}=\ket{\psi}$, and divide by $\braket{\psi|\psi}$ to get
\begin{equation}\label{eq:dynopt}
\bra{\delta \psi}\mathbb{P}_\perp i\partial_t \ket{\psi}=\bra{\delta \psi}\big(\hat{\mathcal{H}}-\braket{\hat{\mathcal{H}}^\dagger}-R\big)\ket{\psi},
\end{equation}
where $\mathbb{P}_\perp=(1-\ket{\psi}\braket{\psi|\psi}^{-1}\bra{\psi})$ is a projector onto the subspace orthogonal to $\ket{\psi}$, and $\braket{\hat{\mathcal{H}}^\dagger}$ is the normalized expectation value.
The variable $R$ is related to the Lagrange multiplier $\lambda$, but will not appear in the final equations.  A more detailed derivation of this equation is presented in appendix \ref{sec:AppB}, which gives more insight into this parameter.

Because of the presence of the projector in equation \eqref{eq:dynopt}, only the perpendicular part of the derivative is constrained by this equation. The parallel part determines the normalization of the wavefunction, which we set by imposing the condition
\begin{equation}\label{eq:perpcond}
\frac{\bra{\psi}i\partial_t\ket{\psi}}{\braket{\psi|\psi}}=\braket{\hat{\mathcal{H}}^\dagger}+R,
\end{equation}
which simplifies equation \eqref{eq:dynopt} to 
\begin{equation}\label{eq:dynvariational}
\bra{\delta \psi}i\partial_t \ket{\psi}=\bra{\delta \psi}\hat{\mathcal{H}}\ket{\psi}.
\end{equation}
This equation is similar to a time-dependent Hartree Fock equation\cite{li2005time}, but adapted to the thermofield formalism for a more general ansatz.  In order to use this equation, we must demand that the variation $\ket{\delta\psi}$ maintains the form of the ansatz for the wavefunction, and so does the derivative.  In the next two sections, we illustrate how this works for a specific choice of truncation and optimization scheme.

\section{Thermal coherent state evolution}

In this section we will develop a specific implementation of the general approach developed previously that might be suitable for quantum chemistry applications.  We will make the following choices:
\begin{itemize}
\item The cumulant expansion is truncated at the first term, imposing the form of a thermal coherent state.  \item The wavefunction will be chosen to be thermofield normal.
\item The basis of excitations will be a set of single particle molecular orbitals satisfying canonical anti-commutation relations. 
\item We will allow arbitrary unitary transformations on the vector space spanned by the operators $\hat{a}_\alpha$ and $\hat{b}_\alpha$, mixing the two Hilbert spaces.
\end{itemize}

The reason we choose to retain the thermofield normal form is that it simplifies calculations, which will be demonstrated, in addition to the fact that it is exact in the non-interacting case.  We will want to find the differential equations that result from using these choices in the variational equation \eqref{eq:dynvariational}.  There are two parameters that will be optimized in the variation, changes in the operator $\hat{T}$ and changes in the basis of excitations.  In general, changes in the basis of excitations will affect the vacuum $\ket{\Omega}$ in addition to the operator $\hat{T}$.  In order to maintain the thermofield normal form, we demand that when the basis is changed, there is a corresponding change to the wavefunction that leaves the operator $\hat{T}$ form invariant.  More explicitly, the combined effect of this transformation on the wavefunction will be 
\begin{equation}
e^{\sum_{\alpha \beta}t_{\alpha\beta}\hat{a}_\alpha^\dagger\hat{b}_\beta}\ket{\Omega}\rightarrow e^{\sum_{\alpha \beta}t_{\alpha\beta}\hat{a}_\alpha^{\prime\dagger}\hat{b}^\prime_\beta}\ket{\Omega^\prime},
\end{equation}
where the prime indicates the changed basis.  Changes in the operator $\hat{T}$ will take the form $t_{\alpha\beta}\rightarrow t_{\alpha\beta}^\prime$, so the two types of transformations are effectively decoupled.

We note that the state remains invariant under the following set of simultaneous transformations
\begin{equation}
\hat{a}^\dagger\rightarrow \hat{a}^\dagger U^\dagger, \ \ \ \hat{b}\rightarrow V\hat{b} , \ \ \ t\rightarrow UtV^\dagger,
\end{equation}
where matrix multiplication is implied, and the matrices are unitary.  In the Hartree-Fock approach this type of freedom is used to diagonalize the matrix of Lagrange multipliers, imposing orthonormality on the basis of excitations\cite{ISI:000252851600016}.  In this case, however, the transformations are infinitesimal so it is possible to directly constrain them to satisfy canonical anti-commutation relations.  For this reason, we will use this freedom to perform a singular value decomposition on the matrix $t_{\alpha\beta}$, which means that this matrix can be chosen to be diagonal, with diagonal elements $t_\alpha$.  

The unitary transformations that mix the Hilbert spaces are included to allow for some correlation in the wavefunction. The generators of these transformations take the form
\begin{align}\label{eq:abtrans}
\begin{bmatrix}
\hat{a} \\
\hat{b}
\end{bmatrix}
\rightarrow
\begin{bmatrix}
\hat{a} \\
\hat{b}
\end{bmatrix}
+i\varepsilon
\begin{bmatrix}
A & B\\
B^\dagger & C^\dagger
\end{bmatrix}
\begin{bmatrix}
\hat{a} \\
\hat{b}
\end{bmatrix}
\end{align}
where $A$ and $C$ are Hermitian.
This is a one parameter group of transformations, whose derivative at $\varepsilon=0$ gives the Lie derivative, which in this case will be denoted $\mathcal{L}_{A,B,C}$.  The vacuum transforms as
\begin{equation}
i\mathcal{L}_{A,B,C}\ket{\Omega}=  \hat{a}^\dagger B \hat{b}\ket{\Omega}.
\end{equation}
which can be checked by acting with any annihilation operator.  Additionally we find
\begin{equation}\label{eq:psivariation}
i\mathcal{L}_{A,B,C}\ket{\psi}=\Big(\hat{a}^\dagger\big(A^\dagger t- tC^\dagger +B-tB^\dagger t\big)\hat{b} +\text{Tr}(B^\dagger t)\Big)\ket{\psi}.
\end{equation}
The second term in this equation changes the normalization of $\ket{\psi}$, which will not be important.  Consider the matrix in the first term
\begin{equation}
M=A^\dagger t-t C^\dagger +B-tB^\dagger t.
\end{equation}
We note that the diagonal can be chosen to be anything, which implies that we can absorb a change in the singular values $t_\alpha$ into a change in the matrix $M$, allowing us to always choose the singular values to be constant.  For simplicity, we choose $t_\alpha=1$, so the wavefunction takes the form of the thermal coherent state 
\begin{equation}
\ket{\psi}=e^{\sum_\alpha \hat{a}^\dagger_\alpha\hat{b}_\alpha}\ket{\Omega}.
\end{equation}
The matrix $M$ then becomes
\begin{equation}
M=A^\dagger-C^\dagger+B-B^\dagger.
\end{equation}
With this form, it is clear that the state is invariant under transformations with $A=C$ and $B=B^\dagger$.  If we choose $A=-C$ and $B=-B^\dagger$, then this makes the first term Hermitian and the second term anti-Hermitian, which decouples the two transformations.  This is interpreted naturally by defining the operators $\hat{a}\pm \hat{b}$, which is developed below.

\section{Simplified basis}\label{sec:complex}

To make better sense of the above results, we define the operators
\begin{equation}\label{eq:zdef}
\hat{v}_\alpha=\frac{1}{\sqrt{2}}(\hat{a}_\alpha-\hat{b}_\alpha ), \ \ \ \hat{w}_\alpha =\frac{1}{\sqrt{2}}(\hat{a}_\alpha+\hat{b}
_\alpha ),
\end{equation}
which will eventually be considered an initial condition.  
These operators satisfy canonical anti-commutation relations, and anti-commute with each other. Also, the state 
\begin{equation}\label{eq:zvacuum}
\ket{\Omega}\propto e^{\sum_\alpha \hat{a}^\dagger_\alpha \hat{b}_\alpha}\ket{0},
\end{equation}
is annihilated by the  operators $\hat{v}_\alpha$ and $\hat{w}_\alpha^\dagger$, so is proportional to the vacuum of this basis.
One can readily check that for the choices $A=C$ and $B=B^\dagger$, the transformation \eqref{eq:abtrans} can be written succinctly as  
\begin{equation}\label{eq:ztrans}
\hat{v}\rightarrow \Big(1+i\varepsilon X\Big)\hat{v}, \ \ \ \hat{w}\rightarrow \Big(1+i\varepsilon \tilde{X}\Big)\hat{w},
\end{equation}
where we have defined $X=A-B$ and $\tilde{X}=A+B$, which are both Hermitian.  

  With this result, it is clear why the vacuum is invariant under these transformations, because they rotate the creation and annihilation operators independently.  For the transformations with $A=-C$ and $B=-B^\dagger$,  we find
\begin{equation}\label{eq:ztrans2}
\hat{v}\rightarrow \hat{v}+i\varepsilon Y\hat{w}, \ \ \ \hat{w}\rightarrow \hat{w}+ i\varepsilon Y^\dagger\hat{v},
\end{equation}
where $Y=A+B$, but this matrix is not Hermitian unless $B=0$.  Under this change, the vacuum  transforms as 
\begin{equation}\label{eq:vactrans}
i\mathcal{L}_Y\ket{\Omega}=\hat{v}^\dagger Y\hat{w}\ket{\Omega},
\end{equation}
which can be checked by acting with annihilation operators at small $\varepsilon$.

These equations are simplified by defining a Hermitian  operator
\begin{equation}\label{eq:effH}
\hat{\mathcal{H}}^c=\hat{v}^\dagger Y \hat{w}+\hat{w}^\dagger Y^\dagger \hat{v}+\hat{v}^\dagger X\hat{v} +\hat{w}^\dagger \tilde{X}\hat{w}.
\end{equation}
The transformations described so far can then be written succinctly as 
\begin{equation}\label{eq:heistrans}
i\mathcal{L}_{\mathcal{H}^c}\ket{\Omega}=\hat{\mathcal{H}}^c\ket{\Omega}, \ \ \ i\mathcal{L}_{\mathcal{H}^c}\hat{A}=[\hat{\mathcal{H}}^c, \hat{A}],
\end{equation}
where $\hat{A}$ refers to $\hat{v}$ or $\hat{w}$.  These equations are naturally interpreted as generating Hamiltonian dynamics in the Schrodinger picture, where the operator $\hat{\mathcal{H}}^c$ can be found from equation   \eqref{eq:dynvariational}. Before developing these dynamics further, we will use the unitary freedom to relate the defined basis to the original basis of operators $\hat{a}_\alpha$.
 
The Hamiltonian $\hat{\mathcal{H}}$ is only a function of the operators in the first copy of the original Hilbert space, denoted $\mathfrak{H}_1^{0}$, so we would like to find a relation between the transformed basis and the original operators. To simplify notation, we will now consider the operators $\hat{a}$ to always lie in $\mathfrak{H}_1^{0}$, taking the operators $\hat{v}$ and $\hat{w}$ to rotate independently of this basis,  meaning that relation \eqref{eq:zdef} no longer holds for $t>0$.  Denote the initial basis of operators by $\hat{a}_0$, and relate the complex basis to the initial basis by the relation
\begin{equation}
\hat{a}_0=M\hat{v}+N\hat{w},
\end{equation} 
for some matrices $M$ and $N$. In appendix \ref{sec:AppC} we prove that these matrices have the same left singular vectors, and there is a unitary change of basis that brings this to the form
\begin{equation}\label{eq:arelation}
\hat{a}=\tilde{\Lambda}\hat{v}+\Lambda\hat{w},
\end{equation} 
where $\Lambda$ and $\tilde{\Lambda}$ are real diagonal matrices with eigenvalues in the range $[0,1]$, related by
\begin{equation}\label{eq:angleinterp}
\tilde{\Lambda}^2+\Lambda^2=1.
\end{equation} 
The operators $\hat{a}$ are now a time dependent basis for the first Hilbert space, unitarily related to the initial basis $\hat{a}_0$. These relations simplify the calculation of physical correlators, which depend only on operators in $\mathfrak{H}_1^{0}$. As an example, we see that the eigenvalues $\lambda_\alpha$ are closely related to the occupation number of the state $\alpha$, as seen by the relation 
\begin{equation}
n_\alpha=\braket{\hat{a}^\dagger_\alpha\hat{a}_\alpha}=\lambda_\alpha^2.
\end{equation} 
In light of equation \eqref{eq:angleinterp}, one could also define a mixing angle $\theta_\alpha$ for which $\lambda_\alpha=\cos(\theta_\alpha)$ and $\tilde{\lambda}_\alpha=\sin(\theta_\alpha)$, but we will continue to use the defined notation.

\section{Approximate dynamics}

To find the equations of motion, we will demand that $\ket{\Omega_c}$ retains the form of the ansatz, implying that it changes according to \eqref{eq:vactrans} 
\begin{equation}
i\partial_t\ket{\Omega}=\hat{v}^\dagger Y \hat{w}\ket{\Omega}.
\end{equation}
 Using this form for the derivative in equation \eqref{eq:dynvariational}, and setting the variation to zero gives 
\begin{equation}\label{eq:Heff}
Y_{\alpha\beta}=\braket{\hat{w}_\beta^\dagger\hat{v}_\alpha\hat{H}}.
\end{equation}
We define the matrix $Y_{\alpha\beta}$ for real time evolution, using \eqref{eq:calH} to find the corresponding imaginary time evolution. Consider now a Hamiltonian of the form \eqref{eq:fullHamiltonian}.    Define time dependent coefficients $H_{\alpha\beta}^{(0)}(t)$ and $w_{\alpha\beta\gamma\delta}(t)$ to expand the Hamiltonian in the basis $\hat{a}=U\hat{a}_0$, 
\begin{equation}\label{eq:unitarychange}
H^{(0)}_{\alpha\beta}(t)=\sum_{\gamma_\delta}U_{\alpha\gamma}H^{(0)}_{\gamma\delta}(0)U_{\delta\beta}^\dagger, \ \ w_{\alpha\beta\gamma\delta}(t)=\sum_{\epsilon\zeta\eta\theta}U_{\alpha\epsilon}U_{\beta\zeta}w_{\epsilon\zeta\eta\theta}(0)U_{\eta\gamma}^\dagger U_{\theta\delta}^\dagger,
\end{equation}  
where the time dependence will be supressed for notational convenience. Equation \eqref{eq:Heff} then becomes
\begin{align}\label{eq:Hcmat}
\frac{Y_{\alpha\beta}}{{\tilde{\lambda}_\alpha \lambda_\beta}}=H^{(0)}_{\alpha\beta}
+\sum_{\gamma}n_\gamma \Big(w_{\alpha\gamma\gamma\beta}-w_{\alpha\gamma\beta\gamma}-w_{\gamma\alpha\gamma\beta}+w_{\gamma\alpha\beta\gamma}\Big),
\end{align}
 This calculation motivates us to define the matrix 
\begin{align}
H^c_{\alpha\beta}=\frac{Y_{\alpha\beta}}{{\tilde{\lambda}_\alpha \lambda_\beta}},
\end{align} 
which is Hermitian.  For imaginary time evolution, the operator $\hat{\mathcal{H}}$ becomes anti-Hermitian, but in the following we will use this definition and equation \eqref{eq:calH} to find the dynamics explicitly. 
  
We can also define the effective Hamiltonian to induce the unitary transformation on the bases $\hat{a}$, $\hat{v}$, and $\hat{w}$ resulting in equation \eqref{eq:arelation}.  The generators of these unitary transformations are derived in appendix \ref{sec:AppC}, which can then be used in equation \eqref{eq:effH}.  The resulting operator is only simplified for real time evolution, given by
\begin{equation}
\hat{\mathcal{H}}^c=
\hat{a}^\dagger H^c \hat{a}, \ \ \ \ \ \ real.
\end{equation} 
In imaginary time the resulting operator is complicated, and not particularly enlightening, instead the explicit equations are derived below.  
For real time evolution, the equations of motion for the vacuum and bases are the same as in equation \eqref{eq:heistrans}, and the occupation numbers are invariant, reflecting the fact that normalization is preserved by unitary time evolution.

In imaginary time, we will again use the parameter $\tau=\beta/2$.  Using the definition \eqref{eq:calH}, and changing variables to $n_\alpha=\lambda_\alpha^2$ we find\footnote{For this derivation we have assumed that the number operator is given by $\hat{N}=\sum_\alpha \hat{a}^\dagger_\alpha \hat{a}_\alpha$, so that these operators are unitarily related to the original creation and annihilation operators.}
\begin{align}\label{eq:aeom1}
\frac{\partial_\tau n_\alpha}{n_\alpha(1-n_\alpha)} = -2 \, H^c_{\alpha\alpha}, \ \ \ \partial_\tau \hat{a}_{\alpha}=\sum_{\beta\neq\alpha}\frac{n_\alpha+n_\beta - 2n_\alpha n_\beta}{n_\alpha-n_\beta}H^c_{\alpha\beta}\hat{a}_\beta,
\end{align}
where the latter equation is only true if the occupancies $n_\alpha$ are non-degenerate.  The degenerate case is explained in appendix \ref{sec:AppC}. These equations can be simplified further by imposing the form of a Fermi-Dirac distribution
\begin{align}\label{eq:occnum}
n_\alpha = \frac{1}{1+e^{\beta(E_\alpha-\mu)}},
\end{align}
which defines an energy $E_\alpha$.  With this definition, equations \eqref{eq:aeom1} become
\begin{align}
&\tau \frac{d E_{\alpha}}{d\tau} =  H^c_{\alpha\alpha}-E_\alpha
\label{eq:aeomE} \\
& \partial_\tau \hat{a}_{\alpha}=\sum_{\beta\neq \alpha}\coth\Big(\tau(E_\beta-E_\alpha)\Big)H^c_{\alpha\beta}\hat{a}_{\beta} ,\label{eq:aeomU}
\end{align}
where we have used a total derivative in the first equation to emphasize the dependence on $n_\alpha$.

These equations give some physical insight into the dynamics. Equation \eqref{eq:aeomE} shows that the energies are driven towards a self consistent field in which the defined energies $E_\alpha$ are close to the energies $H_{\alpha\alpha}$, which can be considered a mean-field energy for the state. Equation \eqref{eq:aeomU} imposes a unitary transformation which tends to diagonalize the effective Hamiltonian.  In the limit of low-temperature, or large $\tau$, the system is driven towards an equilibrium configuration which approximately diagonalizes the Hamiltonian and in which the occupation numbers approach those of the self consistent field, as is the case for the Hartree Fock solution.  Indeed, as $\beta\rightarrow \infty$, the occupation numbers \eqref{eq:occnum} are driven to zero or one, so the state approaches a Slater determinant.  In this case, the effective Hamiltonian reduces to the Fock operator, and the dynamics approaches the Hartree-Fock solution.

\section{Homogeneous electron gas}

We now apply these approximate dynamics to the homogeneous electron gas.  In this case, the Hamiltonian \eqref{eq:fullHamiltonian} consists of the terms
\begin{align}
\hat{H}^{(0)}=\sum_{\mathbf{k}\sigma}\frac{k^2}{2m}\hat{a}^\dagger_{\mathbf{k}\sigma}\hat{a}_{\mathbf{k}\sigma}, \ \ \ \hat{H}^{(1)}=\frac{2\pi e^2}{V}\sum_{\mathbf{k}\mathbf{k}'\sigma\sigma'}\sum_{\mathbf{q}\neq 0}\frac{1}{q^2}\,\hat{a}^\dagger_{\mathbf{k}-\mathbf{q}\sigma}\hat{a}^\dagger_{\mathbf{k}'+\mathbf{q}\sigma'}\hat{a}_{\mathbf{k}'\sigma'}\hat{a}_{\mathbf{k}\sigma},
\end{align} 
where we use units such that $4\pi \varepsilon_0=1$, and $\mathbf{k}$ label the wavevectors of a periodic box.  For this Hamiltonian, equation \eqref{eq:Hcmat} gives
\begin{equation}
H^c_{\mathbf{k}\sigma ,\mathbf{k}'\sigma'}=\delta_{\mathbf{k}\mathbf{k}'}\delta_{\sigma\sigma'}\left(\frac{k^2}{2m}-\frac{4\pi e^2}{V}\sum_{\mathbf{q}\neq 0}\frac{1}{q^2}n_{\mathbf{k}+\mathbf{q}\sigma}\right),
\end{equation}
which shows that the effective Hamiltonian is always diagonal in the Fourier basis, and that the interaction energy is negative due to the exchange energy, as correlation effects are not included well in this approximation.  In the limit of low density this should become an unstable fixed point, because of the tendency towards spin-polarization and Wigner crystallization, making equation \eqref{eq:aeomU} relevant.  We define the occupation energy of interaction by $E^{int}_{\mathbf{k}\sigma}=E_{\mathbf{k}\sigma}-\frac{k^2}{2m}$, which evolves according to
\begin{equation}\label{eq:intdyn}
-\tau\frac{d E^{int}_{\mathbf{k}\sigma}}{d \tau} = E^{int}_{\mathbf{k}\sigma}+\frac{4\pi e^2}{V}\sum_{\mathbf{q}\neq 0}\frac{1}{q^2}n_{\mathbf{k}+\mathbf{q}\sigma}. 
\end{equation}

We now take the infinite volume limit, where $\mathbf{k}$ becomes a continuous variable.  We assume that the density only depends on the magnitude $|\mathbf{k}|$, and omit the dependence on $\sigma$.   The sum in equation \eqref{eq:intdyn} then approaches an integral, defined by
\begin{equation}
\mathcal{I}_{\beta\mu}(k)= \frac{e^2}{2\pi^2}\int  \frac{d^3 q}{q^2}n_{\beta\mu}(|\mathbf{k}+\mathbf{q}|),
\end{equation}
In this integral, there are three relevant distances: $|\mathbf{k}|$, $|\mathbf{q}|$, and $|\mathbf{k}+\mathbf{q}|$.  This situation lends itself to the two-center bipolar coordinate system, related to the upper half-plane by
\begin{equation}
x=\frac{r_1^2-r_2^2}{4a}, \ \ \ \ y=\frac{1}{4a}\sqrt{(4ar_2)^2-(r_2^2-r_1^2+4a^2)^2}.
\end{equation}
In this case, we rotate this coordinate system around the axis connecting the points defining the coordinate system, which introduces an extra angle $\phi$.  The volume element is given by
\begin{equation}
dV=\frac{r_1r_2}{2a}dr_1 d r_2 d \phi.
\end{equation}
Choosing $k=2a$,  $r_1=q$ and $r_2=|k+q|$, we find
\begin{align}\label{eq:nintegral}
\mathcal{I}_{\beta\mu}(k)= \frac{e^2}{\pi}\int_0^\infty dx \log\left(\frac{k+x}{|k-x|}\right)x\,n_{\beta\mu}(x).
\end{align}
In the low temperature limit, when the occupation numbers are filled until the Fermi level, this reduces to a standard result in the Hartree-Fock analysis of the homogeneous electron gas\cite{Ashcroft}.

\section{Conclusion}
We have found a natural way of using thermofield dynamics for imaginary time evolution in the grand canonical ensemble, showing that the approach works for a non-interacting system.  We then applied this analysis to interacting systems, giving rise to a general strategy for approximating the time-dependent Schrodinger equation in a grand canonical ensemble, by introducing an ansatz and optimizing it variationally at every time step to yield approximate equations of motion.  We implemented this strategy for a thermal coherent state, which turns out to be a generalization of time dependent Hartree-Fock theory to fractional occupation numbers.  It will be useful to consider how to include correlation into this ansatz in future work.

\appendix

\section{Canonical transformations}\label{sec:AppA}
The most general canonical transformation that retains the type of statistics can formally written as a unitary transformation on the many-particle Hilbert space\cite{PhysRevB.44.12413}\cite{citeulike:13177435}.  For a system of $N$ particle excitations, the many particle Hilbert space is spanned by $2^N$ basis functions, which are labeled $\chi_i$, where we take $\chi_0$ to be the vaccum. The group $SU(2^N)$ acts on this basis, whose action can be written as a polynomial of Fermi operators.  To do this, define Wenger's matrix, which satisfies $m_{ij}\chi_k=\chi_{i}\delta_{jk}$.  This can be generated by eplicitly constructing $m_{0j}$, then using the relations $m_{i,0}=m^\dagger_{0,i}$ and $m_{i,j}=m_{i,0}m_{0,j}$.  Given a matrix $x\in SU(2^N)$, define the polynomial 
\begin{equation}
P(x)=Tr(xm).
\end{equation}
When acting on the Hilbert space in the usual way, this polynomial induces the unitary transformation $x$.  Also, these polynomials satisfy $P(x)P(y)=P(xy)$ and $P(x^\dagger)=P(x)^\dagger$.  The most general canonical transformation of operators can then be written
\begin{equation}
\hat{a}_{\alpha}\rightarrow P(x)^\dagger \hat{a}_{\alpha} P(x).
\end{equation}
These transformations are probably too general for practical calculation, but this construction shows how large the space of canonical transformations is for a given Hilbert space.

We can also consider canonical transformations from an infinitesimal point of view
\begin{equation}
\hat{a}_\alpha\rightarrow \hat{a}_\alpha+\delta\hat{a}_\alpha,
\end{equation}
For this transformation to be canonical, the variation must satisfy 
\begin{equation}
[\hat{a}_\alpha,\delta \hat{a}_\beta]_\zeta =-[\delta \hat{a}_\alpha, \hat{a}_\beta]_\zeta, \ \ \ [\hat{a}_\alpha,\delta \hat{a}_\beta^\dagger]_\zeta=-[\delta \hat{a}_\alpha,  \hat{a}_\beta^\dagger]_\zeta,
\end{equation}
where we include excitations with Fermionic or Bosonic statistics, defined by $[\hat{a}_\alpha,\hat{a}_\beta]_\zeta = \hat{a}_\alpha\hat{a}_\beta-\zeta\hat{a}_\beta\hat{a}_\alpha$. By inserting the resolution of the identity, we find the variation of the vacuum 
\begin{equation}
\ket{\Omega}\rightarrow \left(1-\sum_\alpha\hat{a}^\dagger_\alpha \frac{1}{\hat{N}_a+1}\delta \hat{a}_\alpha\right)\ket{\Omega},
\end{equation}
where $\hat{N}_a=\sum_\alpha \hat{a}^\dagger_\alpha \hat{a}_\alpha$ is the number operator in this basis. This can also be checked by acting with the transformed annihilation operators.

\section{Incremental optimization}\label{sec:AppB}
Here we derive equation \eqref{eq:dynopt} in a more rigorous way.  First, we define $\ket{\psi_0}=\ket{\psi(t)}$ and $\ket{\psi}=\ket{\psi(t+\delta t)}=(1-i\delta t \hat{\mathcal{H}})\ket{\psi_0}$.  Assume that $\braket{\psi_0|\psi_0}=1$, which can always be chosen.  Define $\ket{\upsilon}=\ket{\psi_0}+\ket{\epsilon}$, where $\ket{\epsilon}$ is small, and similarly $\lambda=1+\epsilon$. With these definitions, and keeping only terms of at most second order in small quantities, equation \eqref{eq:variation1} becomes
\begin{equation}
\bra{\delta\psi_0}\Big(1-\ket{\psi_0}\bra{\psi_0}\Big)i\ket{\epsilon}=\delta t \bra{\delta\psi_0}\left(\hat{\mathcal{H}}-\braket{\hat{\mathcal{H}}^\dagger}-\frac{\epsilon}{\delta t}\right)\ket{\psi_0}.
\end{equation}
If we define $\partial_t \ket{\psi}=\lim_{\delta t\rightarrow 0}\frac{\ket{\epsilon}}{\delta t}$ and $R=\lim_{\delta t\rightarrow 0}\frac{\epsilon}{\delta t}$, this gives equation \eqref{eq:dynopt}.

\section{Relations between unitary transformations}\label{sec:AppC}
To prove equation \eqref{eq:arelation}, we first note that 
\begin{equation}\label{eq:avw0relation}
\hat{a}_0=\frac{1}{\sqrt{2}}\big(\hat{v}_0+\hat{w}_0\big). 
\end{equation} 
The basis at time $t$ is related to the original basis by a unitary transformation of the form
\begin{equation}
\begin{bmatrix}
\hat{v}_0 \\ \hat{w}_0
\end{bmatrix}=
\begin{bmatrix}
A & B \\
C & D
\end{bmatrix}
\begin{bmatrix}
\hat{v} \\ \hat{w}
\end{bmatrix}.
\end{equation}
Using this in equation \eqref{eq:avw0relation} yields
\begin{equation}
\hat{a}_0=\frac{1}{\sqrt{2}}\Big((A+C)\hat{v}+(B+D)\hat{w}\Big). 
\end{equation} 
from which one can read off $M=\frac{1}{\sqrt{2}}(A+C)$ and $N=\frac{1}{\sqrt{2}}(B+D)$.
Because the matrix is unitary, the blocks satisfy the constraints
\begin{align}
AA^\dagger+BB^\dagger=I, \ \ \ CC^\dagger+DD^\dagger=I, \ \ \ AC^\dagger+BD^\dagger=0.\label{eq:blockunitary}
\end{align}

The left singular vectors of a matrix $M$ are the eigenvectors of $M M^\dagger$. Using equations \eqref{eq:blockunitary} we find the relation
\begin{equation}\label{eq:mnrelation}
M M^\dagger=I-N N^\dagger,
\end{equation}
which shows that the matrices $M$ and $N$ have the same left singular vectors.  Therefore, there exist unitary matrices $U$, $V$, and $W$ and diagonal matrices $\Lambda$ and $\tilde{\Lambda}$ that satisfy
\begin{equation}
U\hat{a}_0 =\tilde{\Lambda}V\hat{v}+\Lambda W\hat{w},
\end{equation}
where $U^\dagger$ is the matrix of left singular vectors for $M$ and $N$.
Diagonalizing equation \eqref{eq:mnrelation} with this matrix, we find the relation
\begin{equation}\label{eq:lambdarelation}
\tilde{\Lambda}^2=1-\Lambda^2.
\end{equation} 
 Because the singular values can always be chosen to be real, this also implies that they lie in the interval $[0,1]$. 
 
It is useful to obtain differential equations relating the various matrices defined above.  The change in the operators $v$ and $w$ are given by 
\begin{equation}
\partial_t \hat{v} = iY\hat{w}+iX\hat{v}, \ \ \ \partial_t \hat{w} = iY^\dagger\hat{v}+i\tilde{X}\hat{w}.  
\end{equation}
The unitary transformations implemented above are obtained infinitesimally from equation \eqref{eq:ztrans}.  Define the generator of the unitary transformation on the basis $\hat{a}$ by $J$ so that 
\begin{equation}
\partial_t \hat{a} = iJ\hat{a}.
\end{equation}
where $J$, $X$, and $\tilde{X}$ are Hermitian.  Taking a derivative of equation \eqref{eq:arelation} we find
\begin{equation}\label{eq:Liederivative}
 (\tilde{\Lambda}X-J\tilde{\Lambda}+\Lambda Y^\dagger-i\partial_t \tilde{\Lambda})\hat{v}+(\Lambda\tilde{X}- J \Lambda+\tilde{\Lambda}Y-i\partial_t \Lambda)\hat{w}=0.
\end{equation}
If the eigenvalues of $\Lambda$ are non-degenerate, this equation can be solved by choosing
\begin{align}
J_{\alpha\beta}=\dfrac{\big[\Lambda Y^\dagger \tilde{\Lambda}-\tilde{\Lambda}Y \Lambda\big]_{\alpha\beta}}{\lambda_\alpha^2-\lambda_\beta^2} , \ \ \  X_{\alpha\beta}=\dfrac{\big[\tilde{\Lambda}\Lambda Y^\dagger-Y \Lambda \tilde{\Lambda}\big]_{\alpha\beta}}{\lambda_\alpha^2-\lambda_\beta^2}, \ \ \ \tilde{X}_{\alpha\beta}=\dfrac{\big[Y^\dagger\Lambda \tilde{\Lambda} -\Lambda\tilde{\Lambda}Y \big]_{\alpha\beta}}{\lambda_\alpha^2-\lambda_\beta^2}
\end{align}
for $\alpha\neq\beta$.  For the diagonal elements, the real part can also be canceled by the unitary transformations, and the imaginary part gives
\begin{equation}
\partial_t \tilde{\lambda}_{\alpha}=-\text{Im}(\text{Y}_{\alpha\alpha})\lambda_{\alpha},  \ \ \ \partial_t\lambda_{\alpha}=\text{Im}(\text{Y}_{\alpha\alpha})\tilde{\lambda}_{\alpha}.
\end{equation}

 These equations can also be derived from first order perturbation theory on the matrix $MM^\dagger$ under a small transformation \eqref{eq:ztrans2} to find $J$, and similarly for $X$ and $\tilde{X}$.  Under a small transformation, $MM^\dagger=(\tilde{\Lambda}+i\varepsilon\Lambda Y^\dagger)(\tilde{\Lambda}-i\varepsilon Y\Lambda)$.  In the case of degeneracies one must use degenerate perturbation theory, first diagonalize the perturbing matrix in the degenerate subspace.  In the case of $J$, for example, this means first solving the eigenvalue problem 
\begin{equation}
i\varepsilon(\Lambda Y^\dagger\tilde{\Lambda}-\tilde{\Lambda}Y\Lambda)_{\alpha\beta}u_\beta=(\delta \tilde{\lambda}^2_\alpha)u_\alpha
\end{equation}
in this subspace.

\bibliographystyle{Science}
\bibliography{TCE}

\end{document}